\documentclass[aps,prm,twocolumn,superscriptaddress]{revtex4}
\usepackage{graphicx}
\usepackage{color}

\begin{document}
\title{Quasi-free-standing single-layer WS$_2$ achieved by intercalation}
\author{Sanjoy K. Mahatha}
\email{sanjoy.mahatha@phys.au.dk}
\affiliation{Department of Physics and Astronomy, Interdisciplinary Nanoscience Center (iNANO), Aarhus University, 8000 Aarhus C, Denmark}
\author{Maciej Dendzik}
\altaffiliation{Contributed equally}
\affiliation{Department of Physics and Astronomy, Interdisciplinary Nanoscience Center (iNANO), Aarhus University, 8000 Aarhus C, Denmark}
\affiliation{Department of Physical Chemistry, Fritz-Haber-Institut der Max Planck Gesellschaft, Faradayweg 4-6, 14195 Berlin, Germany}
\author{Charlotte E. Sanders}
\affiliation{Department of Physics and Astronomy, Interdisciplinary Nanoscience Center (iNANO), Aarhus University, 8000 Aarhus C, Denmark}
\author{Matteo Michiardi}
\affiliation{Department of Physics and Astronomy, Interdisciplinary Nanoscience Center (iNANO), Aarhus University, 8000 Aarhus C, Denmark}
\author{Marco Bianchi}
\affiliation{Department of Physics and Astronomy, Interdisciplinary Nanoscience Center (iNANO), Aarhus University, 8000 Aarhus C, Denmark}
\author{Jill A. Miwa}
\affiliation{Department of Physics and Astronomy, Interdisciplinary Nanoscience Center (iNANO), Aarhus University, 8000 Aarhus C, Denmark}
\author{Philip Hofmann}
\email{philip@phys.au.dk}
\affiliation{Department of Physics and Astronomy, Interdisciplinary Nanoscience Center (iNANO), Aarhus University, 8000 Aarhus C, Denmark}

\begin{abstract}
Large-area and high-quality single-layer transition metal dichalcogenides can be synthesized by epitaxial growth on single-crystal substrates. An important advantage of this approach is that the interaction between the single-layer and the substrate can be strong enough to enforce a single crystalline orientation of the layer. On the other hand, the same interaction can  lead to hybridization effects, resulting in the deterioration of the single-layer's native properties. This dilemma can potentially be solved by decoupling the single-layer from the substrate surface after the growth via intercalation of atoms or molecules. Here we show that such a decoupling can indeed be achieved for single-layer WS$_2$ epitaxially grown on Ag(111) by intercalation of Bi atoms. This process leads to a suppression of the single-layer WS$_2$-Ag substrate interaction, yielding an electronic band structure reminiscent of free-standing single-layer WS$_2$. 
\end{abstract}

\maketitle

Single-layer (SL) transition metal dichalcogenides (TMDCs) have recently attracted considerable attention because their electronic structure represents massive Dirac fermions with spin and valley degrees of freedom, which  possess appealing properties for electronic, optical and catalytic applications \cite{Geim:2013aa,Chhowalla:2013aa,Xu:2014ac,Manzeli:2017aa}.  SL TMDCs are produced either via exfoliation from bulk crystals \cite{Novoselov:2005ab,Mak:2010aa, Splendiani:2010aa} or by direct synthesis methods \cite{Koma:1999aa,Schmidt:2014aa,Kang:2015aa}. A particularly promising route for synthesizing large-area, high-quality SL TMDCs is to grow them epitaxially on single-crystal substrates. In the case of a relatively strong TMDC-substrate interaction, typically seen for SL TMDCs grown on single-crystal metal surfaces  \cite{Lauritsen:2007aa, Miwa:2015aa, Dendzik:2015aa}, it is possible to direct the crystalline orientation of the SL with respect to the substrate \cite{Bana:2018aa,Bignardi:2018aa}; a situation that is more difficult to achieve in weakly interacting epitaxial systems, such as SL TMDCs on graphene \cite{Ugeda:2016aa,Miwa:2015ab}. While a strong TMDC-substrate interaction can be advantageous for controlling the growth orientation, a notable drawback of this approach is that the interaction can significantly affect the electronic structure of the SL TMDC due to hybridization between the SL and substrate states \cite{Dendzik:2015aa,Bruix:2016aa,Dendzik:2017ab,Shao:2018aa}. This effect has been studied in great detail for the growth of epitaxial graphene on strongly interacting substrates such as Ni(111) \cite{Varykhalov:2008aa}, Ru(0001) \cite{Sutter:2008aa} and Re(0001)\cite{Papagno:2013aa}, where hybridization between the graphene $\pi$-states and the substrate leads to complete destruction of the Dirac cone. 
 
In the case of epitaxial graphene, the interaction with the substrate can be drastically reduced by the intercalation of small species, such as Au, Ag, oxygen or hydrogen. In the aforementioned cases,  such an intercalation process yields a $\pi$-band dispersion resembling that of free-standing graphene \cite{Varykhalov:2008aa,Sutter:2010aa,Papagno:2013aa}, and even for weakly interacting substrates such as Ir(111) or SiC(0001), intercalation can reduce the graphene-substrate interaction even further \cite{Larciprete:2012aa,Riedl:2009aa}. It has even been shown that the stepwise intercalation of silicon and oxygen can lead to silicon oxide formation under epitaxial graphene, electrically isolating the graphene from the highly conductive substrate on which it is grown \cite{Lizzit:2012aa},  and re-establishing the innate properties of the graphene. Further effects of intercalation can be the generation of substantial electron or hole doping densities \cite{Bostwick:2007aa,Sutter:2010aa,Walter:2011ab,Ulstrup:2014ad}, an increase in the intrinsically very weak spin-orbit interaction  \cite{Marchenko:2012aa,Klimovskikh_2017} and the induction of superconductivity  \cite{Ichinokura_2016}. 

Given this ability to bestow graphene with new properties and to suppress the graphene-substrate interaction by intercalation, it is tempting to assume that epitaxially grown SL TMDCs could equally profit from such an approach. For instance, the intercalation of  ferromagnetic materials could potentially lift the valley degeneracy via exchange coupling, and even a mere change of doping could lead to different charge density wave states \cite{Sanders:2016aa} or superconductivity \cite{Costanzo:2016aa}. Despite these very promising prospects, the fact that bulk TMDCs are generally prone to intercalation \cite{Dresselhaus:1986aa}, and the recent encouraging results of small cation intercalation of SL TMDCs on SiO$_2$ and sapphire substrates \cite{Yu:2017ad}, the intercalation of atoms or molecules in SL TMDC systems has not been fully exploited.

In this paper, we show that the intercalation of Bi atoms  between epitaxial SL WS$_2$ grown on Ag(111) can sufficiently reduce the SL WS$_2$-Ag substrate interaction to restore the band structure of the SL WS$_2$ to the free-standing case. We investigate this process by probing the electronic properties of this material system using angle-resolved photoemission spectroscopy (ARPES) and its structural properties by low energy electron diffraction (LEED) and scanning tunnelling microscopy (STM).

\begin{figure*}[ht]
\includegraphics[width=0.5\textwidth]{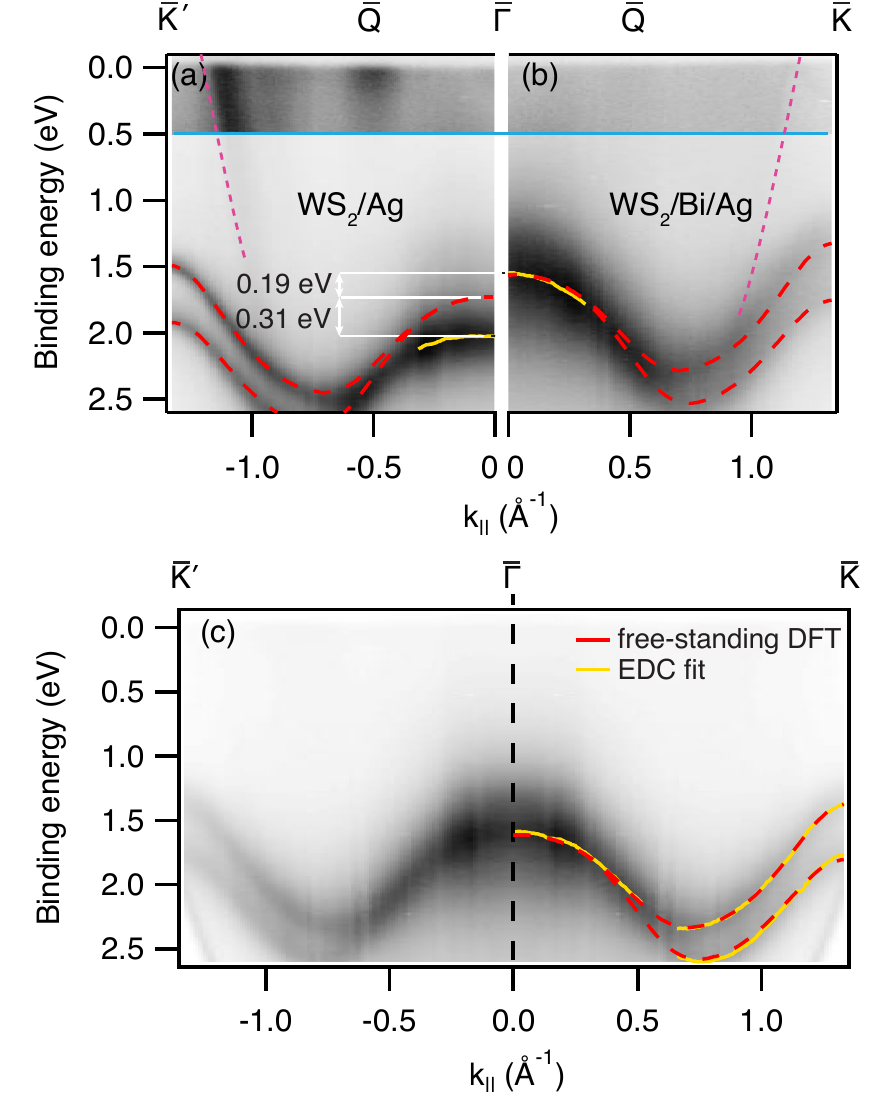}
\caption{Angle-resolved photoemission spectra measured with 30~eV photons of (a) SL WS$_2$ on Ag(111) and (b) SL WS$_2$ on Ag(111) exposed to Bi.  The red dashed lines  overlaid on the experimental data show the calculated free-standing SL WS$_2$ band dispersion from Ref. \cite{Dendzik:2015aa}. In each case, the dashed lines are aligned with the experimental global valence band maximum, situated at $\bar{K}$. The yellow solid lines correspond to fits to the experimental data. The dashed magenta lines denote the limits of the projected bulk band gap. The contrast in the upper part of the spectra, demarcated by the blue solid lines, has been increased to make fainter features visible, in particular the intensity observed near the $\bar{Q}$ point in (a). (c) Same as (b) but acquired with a photon energy of 28~eV.}
\label{fig:1}
\end{figure*}

The Ag(111) single crystal surface was prepared by repeated cycles of noble gas sputtering followed by annealing to 800 K until a sharp (1$\times$1) LEED pattern and an intense Ag(111) surface state was observed in ARPES. SL WS$_2$ was epitaxially grown on Ag(111) following a well-established procedure of W evaporation onto the clean Ag(111) surface in a background pressure of H$_2$S \cite{Lauritsen:2007aa, Miwa:2015aa, Dendzik:2015aa}. Bi was subsequently evaporated from a resistively heated crucible with the substrate held at room temperature, and this was followed by annealing the sample to 520~K. ARPES spectra were acquired at the SGM-3 beamline of ASTRID2 (Aarhus, Denmark) \cite{Hoffmann:2004aa} at a substrate temperature of  30~K. The total energy and angular resolution were better than 30~meV and 0.2$^{\circ}$, respectively.  STM was performed using an Aarhus-type STM integrated in the ultra-high vacuum chamber of SGM-3 beamline end station. STM topography was measured at room temperature using a W tip in constant current mode with the bias applied to the sample.

The valence band dispersion for SL WS$_2$ on Ag(111), measured by means of ARPES using 30\,eV photons, is shown in Fig. \ref{fig:1}(a) and reveals a system rich with hybridization effects. This is best seen when comparing the observed valence band dispersion to a calculation for free-standing SL WS$_2$ (dashed line from Ref. \cite{Dendzik:2015aa}). When the calculation is aligned with the data such that the global valence band maxima at $\bar{K}$ coincide, one finds that the band is severely distorted near $\bar{\Gamma}$, where the energy of the free standing state and the observed band differ by $\approx$ 0.31~eV and the measured state is also rather broad. This broadening and shifting of the band has also been observed for a similar system, i.e. SL MoS$_2$ on Au(111) \cite{Miwa:2015aa,Gronborg:2015aa}, where it is due to strong hybridization of S 3\textit{p$_z$} orbitals, associated with the S atoms of the MoS$_2$ in closest proximity to the Au substrate, with the Au $sp$ states\cite{Bruix:2016aa}. The same explanation holds true here. In contrast, at the $\bar{K}$ point, the SL WS$_2$ bands are sharp and well-defined, and are minimally perturbed by substrate interactions as they lie within the projected band gap of the Ag(111) substrate \cite{Dendzik:2017ab}. Another compelling hybridization feature is situated near the $\bar{Q}$ point and near the Fermi level (in order to enhance the fainter features residing near the Fermi level, the contrast in the upper part of the spectra shown in Fig. \ref{fig:1}(a) and (b) is increased in a binding energy range of 0 -- 0.5~eV).  This photoemission intensity occurs because of a semiconductor-to-metal transition for epitaxial  SL WS$_2$ on Ag(111), caused by a  hybridization  between the conduction bands of the SL WS$_2$ and the Ag substrate  \cite{Dendzik:2017ab}. 

\begin{figure*}[ht]
\includegraphics[width=0.9\textwidth]{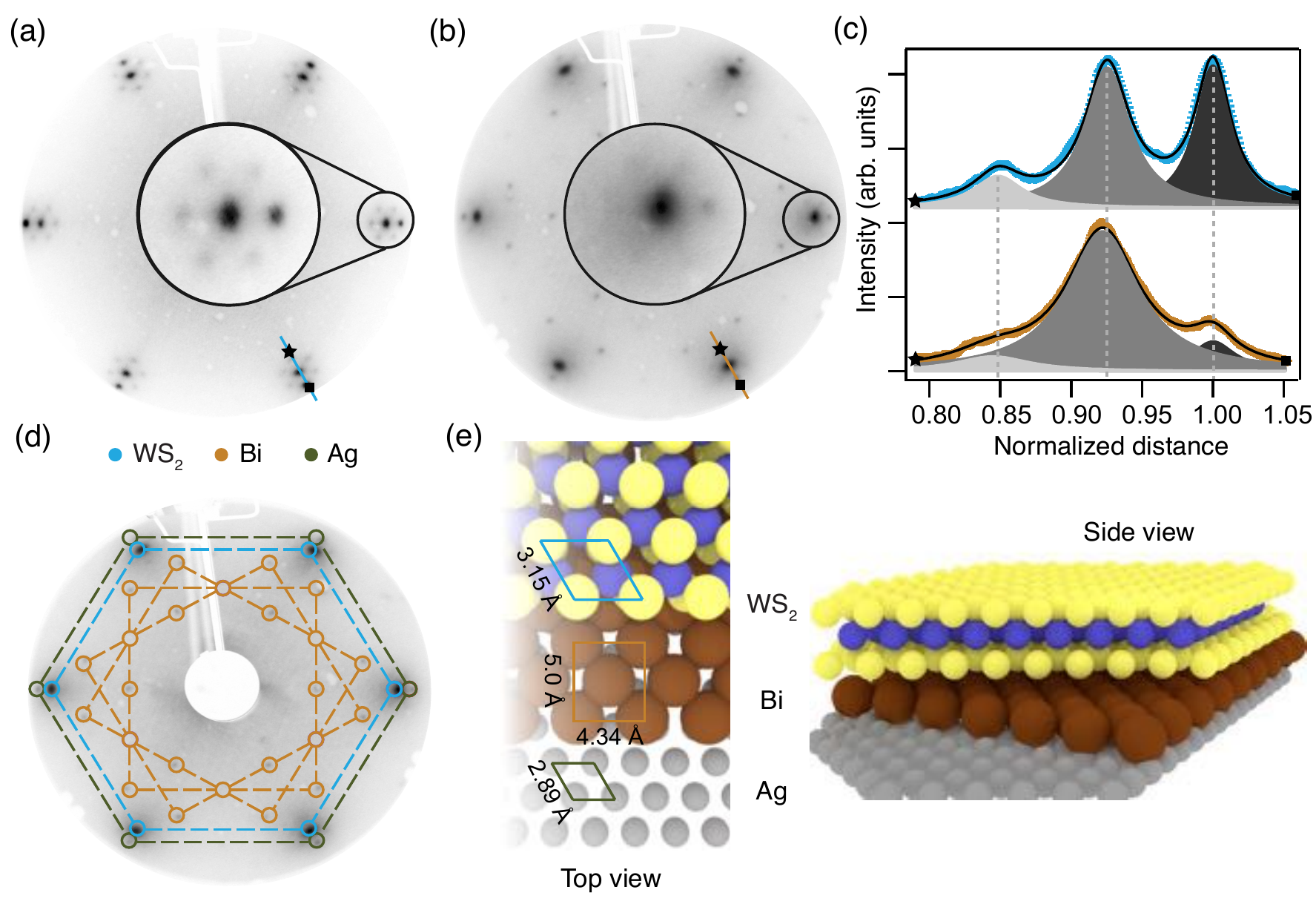}
\caption{Room temperature LEED measurements, acquired with an electron kinetic energy of 64 eV, of SL WS$_2$ on Ag(111) before (a) and after (b) exposure to Bi.  The center insets serve to highlight the diminished intensity of the moir\'{e} satellite spots in (b). (c) Diffracted intensity along radial cuts indicated by the lines of corresponding colour in (a) and (b) and fits to the intensity (black lines), with the fit components indicated in grey. (d) LEED pattern from panel (b) with an indication of the diffraction spots's origin: Ag(111) (green), WS$_2$ (blue) and Bi (brown). (e) A structural model of Bi intercalated SL WS$_2$ on Ag(111) from both top and side perspectives. The unit cells and lattice parameters (extracted from the LEED data) are indicated on the appropriate layer in the top view.}
\label{fig:2}
\end{figure*}

Fig. \ref{fig:1}(b) shows the band structure of the system after exposure to $\approx$ 1 monolayer (ML) amount of Bi, followed by an anneal to approximately 520~K. Compared to the as-grown SL WS$_2$ in  Fig. \ref{fig:1}(a), there are two remarkable differences. First, the valence band in Fig. \ref{fig:1}(b) now closely follows the theoretical dispersion represented by the dashed line, i.e. the distortion of the band structure at $\bar{\Gamma}$ with respect to $\bar{K}$ is removed. Second, there is an absence of photoemission intensity near the Fermi level near $\bar{Q}$.   Both effects are strongly indicative of a significantly weakened SL WS$_2$-substrate interaction.  In addition to the removal of the band structure distortion, Fig. \ref{fig:1}(b) also shows a rigid shift of the entire WS$_2$ valence band towards the Fermi level by 190~meV, i.e. a reduction in n-doping due to the interaction between the SL and the Bi. The upper valence band of SL WS$_2$ after Bi exposure is even better observed when probing the surface with a photon energy of 28~eV and such ARPES results are shown in  Fig. \ref{fig:1}(c). We have used this data set to determine the experimental band position (yellow line) and a comparison to the calculated band dispersion shows excellent quantitative agreement. We also observe an overall broadening of the bands compared to Fig. \ref{fig:1}(a). The reasons for this will be discussed in the end of the paper. We thus conclude that the exposure to Bi leads to a significant weakening of the SL WS$_2$-substrate interaction, restoring the band dispersion expected for free standing SL  WS$_2$ \cite{Zhu:2011ad,Ulstrup:2016ac}. The results depicted in Fig. \ref{fig:1} are reminiscent of studies detailing the transition from graphene that is strongly interacting with its substrate to quasi-free-standing graphene upon intercalation of various small atoms and molecules \cite{Varykhalov:2008aa,Sutter:2010aa,Papagno:2013aa,Riedl:2009aa}. Based on the ARPES data, we propose that SL WS$_2$ on Ag(111), a strongly hybridized system, is transformed to quasi-free-standing SL WS$_2$ via intercalation of Bi atoms.  

From the photoemission data alone we cannot ascertain that the Bi is intercalated between the SL WS$_2$ and the underlying Ag substrate and we therefore study the adsorption by two structural techniques, LEED and STM.  The LEED pattern for SL WS$_2$ on Ag(111) is shown in Fig.~\ref{fig:2}(a). The first order diffraction spots originating from the SL and the underlying substrate both form a hexagonal arrangement (for an assignment of the diffraction spots see Fig.~\ref{fig:2}(d)). The incommensurate atomic lattices of  SL WS$_2$ and  Ag(111) give rise to a  pronounced moir\'{e} pattern that manifests itself as  six satellite spots surrounding each of the main WS$_2$ first order spots in a hexagonal arrangement; one such pattern is magnified in the center inset of Fig.~\ref{fig:2}(a). The structural parameters are determined from a fit of the diffracted intensity shown in the upper panel of Fig.~\ref{fig:2}(c). The blue markers represent the intensity along the blue line in panel (a). It can be fitted with three peaks originating from SL WS$_2$, Ag(111) and the moir\'{e}. From the peak positions, and using the Ag(111) lattice constant as a calibration, the real space lattice constant of SL WS$_2$ is determined to be 3.15(2)~\AA, and the moir\'{e} periodicity is found to be 34.83(3)~\AA. These values are consistent with previous findings for this and similar systems \cite{Sanders:2016aa,Dendzik:2017ab,Gronborg:2015aa}.

\begin{figure*}[ht]
\includegraphics[width=0.9\textwidth]{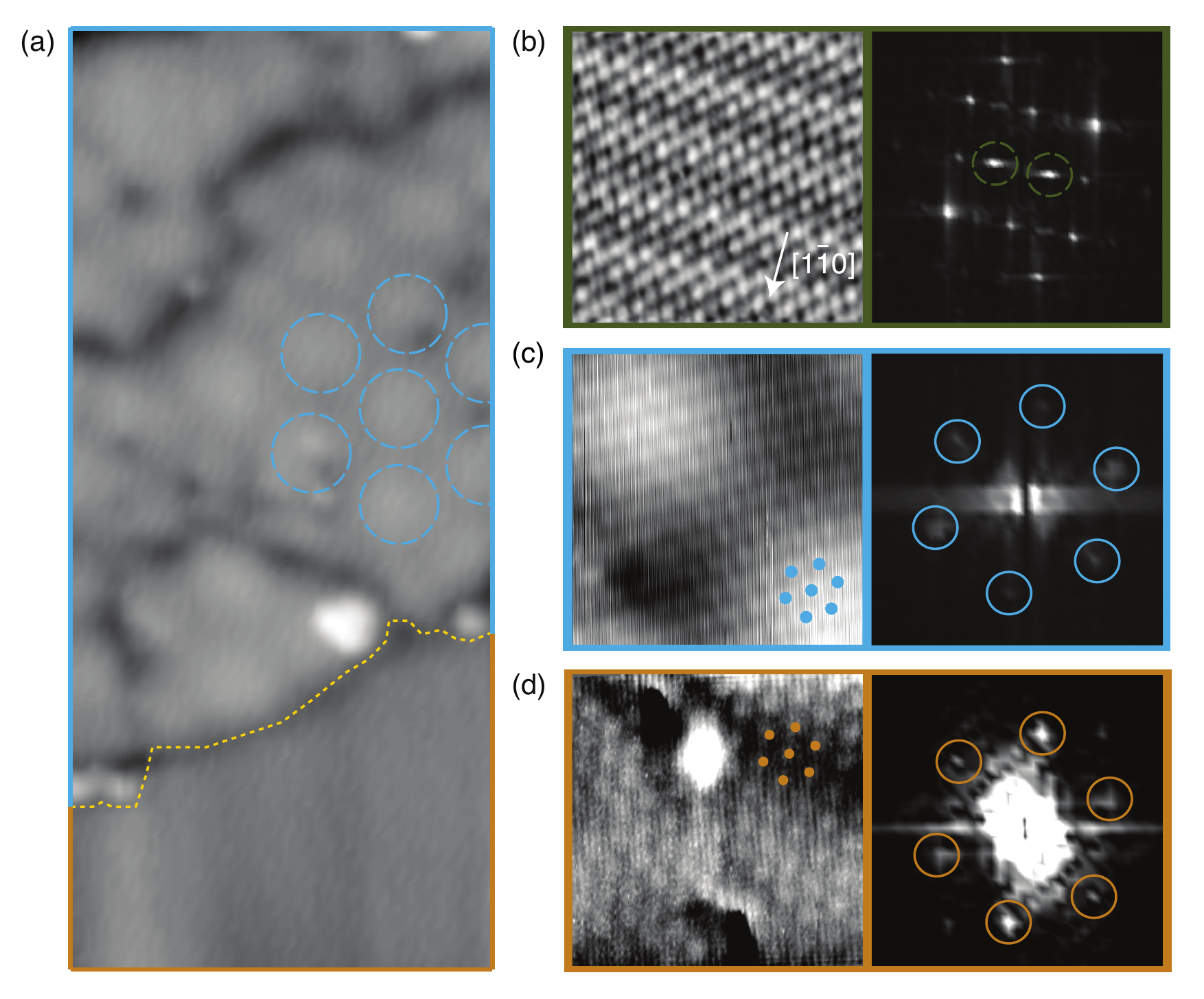}
\caption{(a) STM image showing two different regions of the surface separated by  the  yellow dotted line. The upper region possesses a hexagonal moir\'{e} pattern highlighted by the dashed blue circles, which is consistent with the moir\'{e} expected for SL WS$_2$ on Ag(111). In the lower region the hexagonal moir\'{e} is absent. The additional features, i.e. dark depressions and thin irregular-shaped lines, are attributed to defects and domain boundaries and are typical topographic characteristics of the SL WS$_2$ epitaxially grown on Ag(111). Image parameters: V$_B$ = 1214\,mV, I$_t$ = 0.220\,nA, 300\,{\AA}$\times$140\,{\AA}. (b)  STM image and corresponding FFT showing a rectangular lattice structure consistent with previous observations of the (p$\times$$\sqrt{3}$) Bi adlayer on Ag(111) \cite{Zhang:2011ae}. The extra spots highlighted by the green dashed circles in the FFT pattern arise from the one-dimensional stripes. Image parameters: V$_B$ = 175\,mV, I$_t$ = 1.140\,nA, 50\,{\AA}$\times$50\,{\AA} (STM) and 10\,{\AA}$^{-1}$$\times$10\,{\AA}$^{-1}$ (FFT). (c) Atomic-resolution STM image and FFT of the upper region of panel (a).  Image parameters: V$_B$ = 94 \,mV, I$_t$ = 1.570\,nA, 30\,{\AA}$\times$30\,{\AA} (STM) and 12\,{\AA}$^{-1}$$\times$12\,{\AA}$^{-1}$ (FFT). (d) Atomic-resolution STM image and FFT of the lower region of panel (a).  Image parameters: V$_B$ = 1214\,mV, I$_t$ = 1.110\,nA, 30\,{\AA}$\times$30\,{\AA} (STM) and 12\,{\AA}$^{-1}$$\times$12\,{\AA}$^{-1}$ (FFT).}  
\label{fig:3}
\end{figure*}

After SL WS$_2$ on Ag(111) is exposed to Bi, the diffraction pattern in Fig.~\ref{fig:2}(b) is observed. All the diffraction spots previously seen in Fig.~\ref{fig:2}(a) are still visible. According to the fit of the   diffracted intensity depicted in the lower half of Fig.~\ref{fig:2}(c), the  lattice constants for SL WS$_2$ and the moir\'{e} remain unchanged. However, it is evident that the  intensity of the moir\'{e} spots is significantly weaker (see also the center insets of Fig.~\ref{fig:2}(a) and (b)).  We therefore conclude that the integrity of the SL WS$_2$ atomic structure is not compromised by the presence of Bi, but the loss of intensity in the diffraction spots associated with the moir\'{e}  indicates a weakened SL WS$_2$-Ag substrate interaction.  Note that the multiple-scattering processes inherent in LEED measurements render such a  simplistic interpretations of the diffraction intensities somewhat problematic. We emphasise, however, that the moir\'{e} spots are consistently weaker at all electron energies probed. The finding of a reduced interaction between SL WS$_2$ and the substrate is consistent with the ARPES results in Fig. \ref{fig:1}, supporting the hypothesis of a Bi intercalation-induced transition to free-standing SL WS$_2$.

Several other additional diffraction spots are noticeable in the LEED pattern shown in Fig. \ref{fig:2}(b), which are explained in terms of the Bi adsorption structure. On clean Ag(111),  Bi adopts a rectangular absorption structure with a p$\times$$\sqrt{3}$ ratio, where p$\approx$2,  commensurate in the $[11\bar{2}]$ direction  and incommensurate in the $[1\bar{1}0]$ direction of the Ag substrate \cite{Chen_1993,Zhang:2011ae}.  Following the symmetry of the Ag(111) surface, the structure exists in three equivalent rotational domains. The diffraction spots originating from these three domains are noted by the brown circles and brown dashed lines that form the three rectangles seen in Fig. \ref{fig:2}(d). The (p$\times$$\sqrt{3}$)-Bi/Ag(111) structure is observed for a Bi coverage of $\approx$ 1~ML whereas the much-studied ($\sqrt{3}$$\times$$\sqrt{3}$) surface alloy phase \cite{Ast:2007aa} is formed for 0.33~ML Bi on Ag(111). 

We carried out room temperature STM measurements to shed further light on the structure of the Bi exposed SL WS$_2$  on Ag(111) sample. The STM data in Fig.~\ref{fig:3}(a) reveals an area of the surface with two distinctly different regions;  the superimposed yellow dotted line marks the boundary between them.  The upper region exhibits a moir\'{e} which manifests as bright round protrusions, noted by the dashed blue circles, arranged in a hexagonal pattern with a measured lattice constant of 32(4)~{\AA}. As seen in Fig. \ref{fig:3}(c), the atomically resolved STM image and its corresponding fast Fourier transform (FFT)  show that the atomic lattice in this region of the surface is also hexagonal (filled blue dots) and possesses a lattice constant of 3.2(3)~{\AA}, which is ten times smaller than that of the moir\'{e}.  Both lattice parameters are in agreement with the values calculated from the LEED data (see Fig.~\ref{fig:2}(c)) and with those previously reported for SL WS$_2$ on Ag(111) \cite{Dendzik:2017ab}. In stark contrast, the lower region of the STM image in Fig.~\ref{fig:3}(a) lacks the hexagonal moir\'{e} pattern.  The atomically resolved STM image acquired in this region possesses a hexagonal structure (brown filled dots) with an atomic lattice constant of 3.2(3)~{\AA} (see Fig.~\ref{fig:3}(d)), which notably matches the one seen in the upper region. The equivalent atomic lattice structures for the upper and lower portions of the STM image suggest that both regions of the surface are covered in SL WS$_2$. Previously reported STM data of SL WS$_2$ on Ag(111) only possess regions of SL WS$_2$ that exhibit both the atomic lattice and moir\'{e} \cite{Dendzik:2017ab}. The absence of the  moir\'{e} in the lower part of the STM image is thus indicative of a weakened SL WS$_2$-Ag substrate interaction and is attributed to the intercalation of Bi atoms between the SL WS$_2$ and the Ag(111) surface; a process that does not lead to any significant change of the nascent SL WS$_2$ atomic lattice constant. 

Although not shown in Fig.~\ref{fig:3}(a), approximately 0.2 ML of the surface adopts  a rectangular atomic lattice structure with a one-dimensional striped modulation which can be seen in Fig.~\ref{fig:3}(b). This structure is consistent with the previously reported (p$\times$$\sqrt{3}$)-Bi structure on Ag(111) \cite{Zhang:2011ae}.  The FFT of the image clearly reveals this rectangular pattern, and shows the lattice constants to be 4.9(5)~{\AA} and 4.5(5)~{\AA}, consistent with the LEED data; see the surface unit cell parameters in Fig.~\ref{fig:2}(f).  The extra spots indicated by the green dashed circles in the FFT pattern arise from the one-dimensional stripes, caused by the moir\'{e} forming between the substrate and the Bi structure. Its lattice constant is 9.8(9)~{\AA}, consistent with a modulation of two times the surface unit cell of the (p$\times$$\sqrt{3}$)-Bi along the shorter axis (Ag$[1\bar{1}0]$ direction).  We ascribe the formation of this structure to Bi adsorption on the parts of the surface not previously covered by SL WS$_2$. Bi atoms could also adsorb on top of the  SL WS$_2$ but there is little topographic evidence of this occurring. Only some bright disordered features covering \textless0.04 ML of the surface are observed, which are presumed to be small Bi clusters. 

Finally, we return to the question why the spin-split bands around $\bar{K}$ are broader for the intercalated SL WS$_2$ system than for the pristine WS$_2$ on Ag(111). There are several contributions to this: (1) The first cause of the broadening is disorder in the system. In the STM data, we observe defects on the surface that we interpret as non-intercalated Bi. These and other defects act as scattering centres, reducing the lifetime of the photohole (and thus increasing the linewidth), for all the WS$_2$ states observed in ARPES. An experimental indication of the disorder's significance is that the broadening can be reduced by annealing the sample. (2) STM also shows the presence of non-intercalated WS$_2$ regions. Since these regions give rise to a shifted band structure (see Fig. \ref{fig:1}(a) and (b)), their presence does also lead to broadening of the features in ARPES. This effect is probably not too important because, due to the substantial size of the shift, a significant amount of non-intercalated regions would stand out clearly in the ARPES data. (3) As already pointed out, the reason for the sharp bands near $\bar{K}$ in the non-intercalated system is that these states are placed in a projected band gap of the substrate band structure, i.e. there are no Ag(111) states at the same energy and $k$ present for hybridization. In the case of the (p$\times$$\sqrt{3}$) structure of adsorbed Bi, this is no longer the case because of Bi-derived bands in the same spectral region. Even if we do not observe any pronounced signs of hybridization (such as splittings or a change in the dispersion), the interaction between the  WS$_2$ and Bi states can be sufficiently strong to induce a lifetime broadening. (4) Finally, the moir\'{e} for the WS$_2$/Bi/Ag(111) surface system contains many short reciprocal lattice vectors that can lead to weak replica bands, simultaneously existing with the original WS$_2$ bands and giving the appearance of broadening \cite{Dendzik:2016aa}. To make matters even more complicated, the rectangular Bi lattice exists in three rotational domains under the WS$_2$ lattice. Indeed, the small reciprocal lattice vectors can also lead to replicas of the projected bulk band structure, closing the gap around $\bar{K}$. Broadening of the bands is thus to be expected for this system, even when no structural defects are present.

In conclusion, we have demonstrated that intercalation can be used to restore the free-standing properties of SL TMDCs that are epitaxially grown on metal surfaces. This opens the possibility to exploit the advantages of epitaxial growth (high quality and orientation of the SL) while avoiding the most prominent disadvantage of this method (strong SL-substrate interaction). Intercalation-based decoupling and modification of properties is already widely used for epitaxial graphene and we expect to see similar applications for SL TMDCs. Indeed, given the wide variety of physical properties in SL TMDCs, there are even more properties that could be tuned, such as charge density wave and superconducting transitions via doping, or the valley degeneracy via magnetic exchange with the intercalated species.

This work was supported by the Danish Council for Independent Research, Natural Sciences under the Sapere
Aude program (Grants No. DFF-4002-00029 and DFF-6108-00409) and by VILLUM FONDEN via the Centre
of Excellence for Dirac Materials (Grant No. 11744) and the Aarhus University Research Foundation.

\end{document}